\documentclass[%
 reprint,
 superscriptaddress,
 amsmath,amssymb,
 aps,
]{revtex4-2}

\usepackage{graphicx}
\usepackage{dcolumn}
\usepackage{bm}
\usepackage{appendix}
\usepackage{hyperref}

\newcommand{\sodium}{\ensuremath{{}^{23}}Na}
\newcommand{\iodine}{\ensuremath{{}^{127}}I}
\newcommand{\xenon}{\ensuremath{{}^{127}}Xe}
\newcommand{\lead}{\ensuremath{{}^{208}}Pb}
\newcommand{\iron}{\ensuremath{{}^{56}}Fe}
\newcommand{\cevns}{CE\ensuremath{\nu}NS}
\newcommand{\naive}{NaI\ensuremath{\nu}E}
\newcommand{\naivete}{NaI\ensuremath{\nu}ETe}
\newcommand{\piDAR}{\ensuremath{\pi}-DAR}
\newcommand{\nueICC}{\ensuremath{\nu_e}CC-\ensuremath{{}^{127}}I}

\begin{document}


\title{Measurement of Electron-Neutrino Charged-Current Cross Sections on ${}^{127}$I with the COHERENT  NaI$\nu$E Detector}


\date{\today}

\newcommand{\Dukedesc}{\affiliation{Department of Physics, Duke University, Durham, NC, 27708, USA}}
\newcommand{\TUNLdesc}{\affiliation{Triangle Universities Nuclear Laboratory, Durham, NC, 27708, USA}}
\newcommand{\UTKdesc}{\affiliation{Department of Physics and Astronomy, University of Tennessee, Knoxville, TN, 37996, USA}}
\newcommand{\Mephidesc}{\affiliation{National Research Nuclear University MEPhI (Moscow Engineering Physics Institute), Moscow, 115409, Russian Federation}}
\newcommand{\ITEPnewadesc}{\affiliation{National Research Center  ``Kurchatov Institute'' , Moscow, 123182, Russian Federation }}
\newcommand{\USDdesc}{\affiliation{Department of Physics, University of South Dakota, Vermillion, SD, 57069, USA}}
\newcommand{\NCCUdesc}{\affiliation{Department of Mathematics and Physics, North Carolina Central University, Durham, NC, 27707, USA}}
\newcommand{\NCSUdesc}{\affiliation{Department of Physics, North Carolina State University, Raleigh, NC, 27695, USA}}
\newcommand{\Sandiadesc}{\affiliation{Sandia National Laboratories, Livermore, CA, 94550, USA}}
\newcommand{\ORNLdesc}{\affiliation{Oak Ridge National Laboratory, Oak Ridge, TN, 37831, USA}}
\newcommand{\UWdesc}{\affiliation{Center for Experimental Nuclear Physics and Astrophysics \& Department of Physics, University of Washington, Seattle, WA, 98195, USA}}
\newcommand{\LANLdesc}{\affiliation{Los Alamos National Laboratory, Los Alamos, NM, 87545, USA}}
\newcommand{\Laurentiandesc}{\affiliation{Department of Physics, Laurentian University, Sudbury, Ontario, P3E 2C6, Canada}}
\newcommand{\CMUdesc}{\affiliation{Department of Physics, Carnegie Mellon University, Pittsburgh, PA, 15213, USA}}
\newcommand{\IUdesc}{\affiliation{Department of Physics, Indiana University, Bloomington, IN, 47405, USA}}
\newcommand{\VTdesc}{\affiliation{Center for Neutrino Physics, Virginia Tech, Blacksburg, VA, 24061, USA}}
\newcommand{\NCSUnucengdesc}{\affiliation{Department of Nuclear Engineering, North Carolina State University, Raleigh, NC, 27695, USA}}
\newcommand{\WJCdesc}{\affiliation{Washington \& Jefferson College, Washington, PA, 15301, USA}}
\newcommand{\UFdesc}{\affiliation{Department of Physics, University of Florida, Gainesville, FL, 32611, USA}}
\newcommand{\Concorddesc}{\affiliation{Department of Physical and Environmental Sciences, Concord University, Athens, WV, 24712, USA}}
\newcommand{\SLACdesc}{\affiliation{SLAC National Accelerator Laboratory, Menlo Park, CA, 94025, USA}}
\newcommand{\Tuftsdesc}{\affiliation{Department of Physics and Astronomy, Tufts University, Medford, MA, 02155, USA}}
\newcommand{\SNUdesc}{\affiliation{Department of Physics and Astronomy, Seoul National University, Seoul, 08826, Korea}}
\author{P.~An}\email{peibo.an@alumni.duke.edu}\Dukedesc\TUNLdesc
\author{C.~Awe}\Dukedesc\TUNLdesc
\author{P.S.~Barbeau}\Dukedesc\TUNLdesc
\author{B.~Becker}\UTKdesc
\author{V.~Belov}\Mephidesc\ITEPnewadesc
\author{I.~Bernardi}\UTKdesc
\author{C.~Bock}\USDdesc
\author{A.~Bolozdynya}\Mephidesc
\author{R.~Bouabid}\Dukedesc\TUNLdesc
\author{A.~Brown}\NCCUdesc\TUNLdesc
\author{J.~Browning}\NCSUdesc
\author{B.~Cabrera-Palmer}\Sandiadesc
\author{M.~Cervantes}\Dukedesc
\author{E.~Conley}\Dukedesc
\author{J.~Daughhetee}\ORNLdesc
\author{J.~Detwiler}\UWdesc
\author{K.~Ding}\USDdesc
\author{M.R.~Durand}\UWdesc
\author{Y.~Efremenko}\UTKdesc\ORNLdesc
\author{S.R.~Elliott}\LANLdesc
\author{L.~Fabris}\ORNLdesc
\author{M.~Febbraro}\ORNLdesc
\author{A.~Gallo Rosso}\Laurentiandesc
\author{A.~Galindo-Uribarri}\ORNLdesc\UTKdesc
\author{A.C.~Germer}\CMUdesc
\author{M.P.~Green}\TUNLdesc\ORNLdesc\NCSUdesc
\author{J.~Hakenm\"uller}\Dukedesc
\author{M.R.~Heath}\ORNLdesc
\author{S.~Hedges}\email{hedges3@llnl.gov}\altaffiliation{Also at: Lawrence Livermore National Laboratory, Livermore, CA, 94550, USA}\Dukedesc\TUNLdesc
\author{M.~Hughes}\IUdesc
\author{B.A.~Johnson}\IUdesc
\author{T.~Johnson}\Dukedesc\TUNLdesc
\author{A.~Khromov}\Mephidesc
\author{A.~Konovalov}\altaffiliation{Also at: Lebedev Physical Institute of the Russian Academy of Sciences, Moscow, 119991, Russian Federation}\Mephidesc
\author{E.~Kozlova}\Mephidesc
\author{A.~Kumpan}\Mephidesc
\author{O.~Kyzylova}\VTdesc
\author{L.~Li}\Dukedesc\TUNLdesc
\author{J.M.~Link}\VTdesc
\author{J.~Liu}\USDdesc
\author{M.~Mahoney}\CMUdesc
\author{A.~Major}\Dukedesc
\author{K.~Mann}\NCSUdesc
\author{D.M.~Markoff}\NCCUdesc\TUNLdesc
\author{J.~Mastroberti}\IUdesc
\author{J.~Mattingly}\NCSUnucengdesc
\author{P.E.~Mueller}\ORNLdesc
\author{J.~Newby}\ORNLdesc
\author{D.S.~Parno}\CMUdesc
\author{S.I.~Penttila}\ORNLdesc
\author{D.~Pershey}\Dukedesc
\author{C.G.~Prior}\Dukedesc\TUNLdesc
\author{R.~Rapp}\WJCdesc
\author{H.~Ray}\UFdesc
\author{J.~Raybern}\Dukedesc
\author{O.~Razuvaeva}\Mephidesc\ITEPnewadesc
\author{D.~Reyna}\Sandiadesc
\author{G.C.~Rich}\TUNLdesc
\author{J.~Ross}\NCCUdesc\TUNLdesc
\author{D.~Rudik}\altaffiliation{Also at: University of Naples Federico II, Naples, 80138, Italy}\Mephidesc
\author{J.~Runge}\Dukedesc\TUNLdesc
\author{D.J.~Salvat}\IUdesc
\author{J.~Sander}\USDdesc
\author{K.~Scholberg}\Dukedesc
\author{A.~Shakirov}\Mephidesc
\author{G.~Simakov}\Mephidesc\ITEPnewadesc
\author{G.~Sinev}\altaffiliation{Also at: South Dakota School of Mines and Technology, Rapid City, SD, 57701, USA}\Dukedesc
\author{C.~Skuse}\CMUdesc
\author{W.M.~Snow}\IUdesc
\author{V.~Sosnovtsev}\Mephidesc
\author{T.~Subedi}\VTdesc\Concorddesc
\author{B.~Suh}\IUdesc
\author{R.~Tayloe}\IUdesc
\author{K.~Tellez-Giron-Flores}\VTdesc
\author{Y.-T.~Tsai}\SLACdesc
\author{E.~Ujah}\NCCUdesc\TUNLdesc
\author{J.~Vanderwerp}\IUdesc
\author{E.E.~van Nieuwenhuizen}\Dukedesc\TUNLdesc
\author{R.L.~Varner}\ORNLdesc
\author{C.J.~Virtue}\Laurentiandesc
\author{G.~Visser}\IUdesc
\author{K.~Walkup}\VTdesc
\author{E.M.~Ward}\UTKdesc
\author{T.~Wongjirad}\Tuftsdesc
\author{J.~Yoo}\SNUdesc
\author{C.-H.~Yu}\ORNLdesc
\author{A.~Zawada}\TUNLdesc
\author{J.~Zettlemoyer}\altaffiliation{Also at: Fermi National Accelerator Laboratory, Batavia, IL, 60510, USA}\IUdesc
\author{A.~Zderic}\UWdesc


\begin{abstract}
Using an 185-kg NaI[Tl] array, COHERENT has measured the inclusive electron-neutrino charged-current cross section on \iodine{}
with pion decay-at-rest neutrinos produced by the Spallation Neutron Source at Oak Ridge National Laboratory. Iodine is one the heaviest targets for which low-energy ($\leq 50$\,MeV) inelastic neutrino-nucleus processes have been measured, and this is the first measurement of its inclusive cross section. After a five-year detector exposure, COHERENT reports a flux-averaged cross section for electron neutrinos of $9.2^{+2.1}_{-1.8}\times10^{-40}$~cm$^2$. This corresponds to a value that is $\sim$41\% lower than predicted using the MARLEY event generator with a measured Gamow-Teller strength distribution. In addition, the observed visible spectrum from charged-current scattering on \iodine{} has been measured between 10 and 55 MeV, and the exclusive zero-neutron and one-or-more-neutron emission cross sections are measured to be $5.2_{-3.1}^{+3.4} \times 10^{-40}$ and $2.2_{-2.2}^{+3.5} \times 10^{-40}\mbox{ cm}^2$, respectively.
\end{abstract}

\maketitle

\textit{Introduction.}
There are little existing experimental data on low-energy inelastic neutrino-nucleus scattering. For terrestrial-based neutrinos with energy less than 100\,MeV, measurements exist for only seven nuclear targets~\cite{formaggio2012,maschuw1998,COHERENT:2022eoh}. Despite the dearth of experimental data, there are motivations for the study of these interactions in detecting solar and supernova neutrinos~\cite{haxton1988}, improving our understanding of weak interactions with the nucleus~\cite{engel1994}, and quantifying backgrounds for neutrino-scattering experiments.

Inelastic charged-current (CC) neutrino interactions on \iodine{} (\nueICC{}), in particular, have generated interest for solar and supernova neutrino detection. The low $Q$ value (662.3\,keV) for the \nueICC{} interaction, along with the large predicted cross section and long half-life of the resulting \xenon{} nucleus, make iodine a promising target for radiochemical neutrino detection. Recent calculations~\cite{lutostansky2021} have shown that, by measuring the fraction of \nueICC{} events that emit a neutron, an \iodine{} solar neutrino detector can provide information on the fluxes of different types of solar neutrinos.

In existing calculations, there are large variations of the pion decay-at-rest (\piDAR{}) flux-averaged \nueICC{} cross section~\cite{engel1994,kosmas1996,mintz2000,athar2006}. One factor impacting predictions is the weak axial-vector coupling constant, $g_A$. By measuring exclusive cross sections to specific multipoles in the resulting \xenon{} nucleus, it may be possible to learn about the quenching of $g_A$~\cite{engel1994}. Neutrino-nucleus interactions at \piDAR{} sources allow the study of $g_A$ in a weak process at larger momentum transfer ($Q\sim10$s of MeV$/c$) than is achievable through $\beta$-decay experiments ($Q\sim1$\,MeV$/c$). The dependence of $g_A$ on momentum transfer has a large impact on neutrinoless double beta-decay ($0\nu\beta\beta$) experiments, where the $0\nu\beta\beta$ half-life depends on $g_A$ to the fourth power. A review of the $g_A$ quenching problem and its impact on $0\nu\beta\beta$ matrix elements can be found in Ref.~\cite{engel2017}. 

In searches for coherent elastic neutrino-nucleus scattering (\cevns{}), inelastic neutrino-nucleus interactions occurring in the active detector volume or surrounding shielding can form a background~\cite{akimov2017}. Neutrino interactions can produce excited nuclear states that deexcite by emitting neutrons. Neutrino-induced neutrons (NINs) can produce keV-scale nuclear recoils that mimic the \cevns{} signal. Additionally, NINs are one of the only backgrounds that can follow the timing distribution of neutrinos produced by pulsed spallation neutron sources. COHERENT plans to measure \cevns{} on \sodium{} using a tonne-scale NaI[Tl] scintillator array. NINs from \nueICC{} can form a background for this search. While the \cevns{} cross section on \sodium{} is expected to be larger than the \nueICC{} cross section, the exclusive \nueICC{} channel leading to neutron emission on iodine has never been measured. A recent search for NINs on \lead{} observed a cross section substantially lower than existing theoretical predictions, although the source of this suppression is unknown~\cite{COHERENT:2022eoh}.

The exclusive \nueICC{} channel to bound states of \xenon{} (referred to as $0n$) was measured for \piDAR{} neutrinos by E-1213 at the Los Alamos Meson Physics Facility (LAMPF)~\cite{Distel:2002ch}. By extracting \xenon{} produced in a 1540-kg NaI solution and counting its decay, a flux-averaged exclusive cross section of $[2.84\pm0.91\mbox{ (stat)}\pm0.25\mbox{ (syst)}]\times10^{-40}$~cm$^2$ was reported. The radiochemical approach was insensitive to CC interactions leading to the emission of one or more neutrons ($\geq1n$), while the majority of $\nu_e$ emitted at a \piDAR{} source have energies above the neutron emission threshold of 7.246\,MeV~\cite{lutostansky2021}. Additionally, radiochemical approaches are unable to measure the energy dependence of the \nueICC{} cross section. 

The NaI Neutrino Experiment (\naive{}) was deployed by the COHERENT Collaboration to the Spallation Neutron Source (SNS) at Oak Ridge National Laboratory (ORNL) to measure the inclusive \nueICC{} cross section. Details on the detector, calibrations, signal predictions, and results from a five-year search are presented here.

\begin{figure}
  \includegraphics[width=0.29\textwidth]{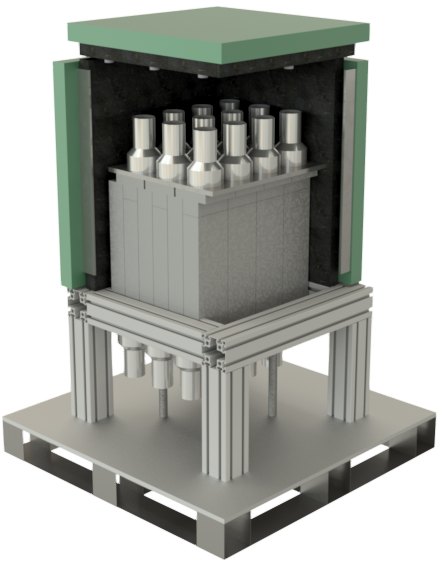}
  \caption{Cutaway view of the 24-crystal \naive{} detector. Muon veto panels are depicted in green and steel shielding in dark gray.}
  \label{fig:schematic}
\end{figure}

\textit{Experimental description.}
The \naive{} detector consists of 24 2'' $\times$ 4'' $\times$ 16'' NaI[Tl] scintillator crystals, each with a mass of $\sim$7.7\,kg, enveloped in thin aluminum shielding. The crystals are arranged in a 4 $\times$ 6 array, oriented vertically, as depicted in Fig.~\ref{fig:schematic}. Each crystal is equipped with a ten-stage 3.5” diameter Burle S83013 photomultiplier tube (PMT). Two-inch plastic scintillator muon veto panels tag muon backgrounds, and 1.5'' of A36 steel rests between the NaI[Tl] crystals and vetoes to avoid vetoing the CC signal. The side (top) veto panels are equipped with two (four) ET-9078B PMTs. The detector began operating at the SNS in its current shielding configuration in 2017.

The detector is located in a basement hallway at the SNS target station, 18.7\,m from the SNS mercury target, where it is exposed to an intense flux of \piDAR{} neutrinos ($\sim5.4 \times 10^7 \mbox{ cm}^{-2} \mbox{s}^{-1}$ at 18.7\,m)~\cite{fluxPaper}. At the SNS, $\nu_e$ neutrinos are the only flavor that can undergo CC interactions, as the muon-neutrino energy is too low to produce muons. The timing of the $\nu_e$ flux is determined by the 350\,ns FWHM proton-on-target (POT) pulse convolved with the 26\,ns mean lifetime of the pion and the 2.2\,$\mu$s mean lifetime of the muon. The maximum energy of the produced $\nu_e$ is $\sim$52.8\,MeV. Additional details on neutrino production at the SNS can be found in Ref.~\cite{fluxPaper}. 

The 24 NaI[Tl] PMTs, 12 muon veto PMTs, and two timing channels from the SNS are read out using five Struck SIS3302 digitizers [eight-channels, 100\,MHz, 16-bit analog-to-digital coverter (ADC)]. Each channel triggers independently using a moving-average trapezoidal trigger. The NaI[Tl] channel trigger thresholds range from 500 to 900\,keV. To reduce the amount of data generated, digitizers store the integrated PMT signal in eight 1.25-$\mu$s windows around the triggering pulse. The first two windows record the baseline level, and the following three integrate pulse signal to define an energy quantity. The digitizer additionally records the peak ADC value of the pulse, the location of the peak within the 10-$\mu$s trace, and whether there is pileup from additional triggers. In analysis, NaI[Tl] signals are correlated with veto and POT signals. One malfunctioning NaI[Tl] crystal was removed from analysis but remained \textit{in-situ} and was simulated.

Energy depositions from multiple NaI[Tl] detectors are grouped to form events if they occur within a 400-ns time window. Long coincidence windows are not atypical for other large NaI[Tl] arrays~\cite{amare2019,lee2023}.


NaI[Tl] scintillator events are associated with cosmic muon events if they occur within a [-6\,$\mu$s,+20\,$\mu$s] window around a muon veto PMT signal. The thresholds for the muon vetoes range from 300 to 700 keV.

Data occurring within a [-2\,$\mu$s,+20\,$\mu$s] window of the POT pulse were blinded to avoid biasing cuts. Health checks removed periods of operation with irregular SNS beam operations or detector electronics issues.

\textit{Detector calibration.}
While NaI[Tl] scintillators are known to exhibit a nonlinear light yield below 1\,MeV~\cite{aitken1967}, the high-energy light yield is fairly linear~\cite{GARDNER1999189}. Each NaI[Tl] detector was calibrated in two steps, first using peaks from low-energy gamma backgrounds, followed by high-energy Michel positrons to account for nonlinearities in the light yield and PMT response. The low-energy calibrations use events from ${}^{40}$K (${}^{208}$Tl) at 1.461 (2.615)\,MeV to estimate the energy scale and track gain changes that occur due to PMT aging and temperature fluctuations. The higher-energy calibrations identify Michel positron candidates by searching to 10-to-55\,MeV depositions following a tagged muon-veto event and is applied to ensure the detector energy response is correct in the energy region of interest (ROI) relevant for CC events in this analysis. Additional details on the calibrations can be found in the Supplemental Material~\cite{naiveSupMat}.



To avoid threshold effects, a software cut removes energy depositions in any crystal below 1\,MeV so that the trigger efficiency is $\sim100\%$.  A second cut removes events with an energy deposition in a single crystal greater than 55\,MeV, as these are outside the \nueICC{} ROI.

\textit{Simulation and signal prediction.}\label{sect:sim} All CC event generation was done with MARLEY~\cite{gardiner2021simulating}. Although designed for CC interactions on ${}^{40}$Ar~\cite{gardiner2021_2}, MARLEY has been adapted for use with other nuclei. MARLEY simulates the allowed component of neutrino-nucleus interactions, relying on provided distributions of the Gamow-Teller (GT$^{-}$) and Fermi (F) strengths. For iodine, $B(\mbox{GT}^-)$ data are obtained from the charge-exchange experiment in Ref.~\cite{I_react}. Similarly, $B(\mbox{GT}^-)$ data in references~\cite{Fe_react,rapaport1983} were used for \sodium{} CC events originating in the NaI[Tl] detectors and \iron{} CC events originating in the shielding. An extended discussion of MARLEY's predictions for \iodine{} can be found in~\cite{naiveSupMat}. The predicted flux-averaged \nueICC{} cross section from MARLEY is $22.5_{-6.5}^{+1.2}\times10^{-40}$~cm$^2$, with uncertainties originating from those provided in the measured GT strength distribution. 

One of the key signatures of CC interactions at \piDAR{} sources is the delayed timing of electron neutrinos resulting from the 2.2\,$\mu$s muon lifetime. This timing distribution is not affected by uncertainties associated with forbidden transitions or $g_A$ quenching, so these factors predominantly affect the predicted spectral shape from MARLEY. The amplitude of the \nueICC{} component is allowed to float in our fits. Backgrounds from neutrino-electron scattering were ignored as their expected rate is only 1.2$\%$ of the expected \nueICC{} signal.

\begin{figure}[!bt]
\centering
\includegraphics[width=0.49\textwidth]{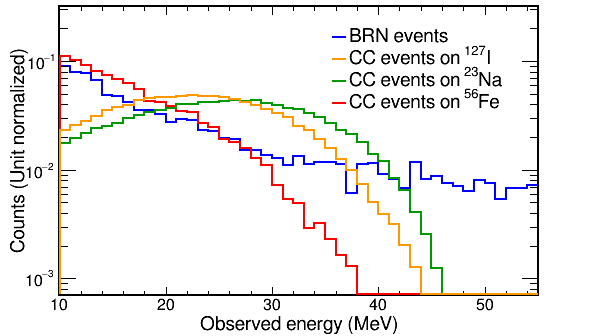}
\caption{Simulated visible energy spectra of \iodine{} CC (orange line), \sodium{} CC (green line), \iron{} CC (red line), and beam-related neutron signals (blue line), after applying cuts. Amplitudes of each component are normalized to unity.}
\label{fig:Sim_Energy}
\end{figure}

A GEANT4~\cite{geant} simulation was used to simulate detector response Michel positrons, beam-related neutrons (BRNs), and CC events on \iodine{}, \sodium{}, and \iron{}. Simulations were postprocessed with cuts designed to match those applied to detector data. Energy smearing was applied to simulated events using the measured energy resolution from the detector calibrations. While nuclear recoils from CC interactions and neutron interactions are below detector thresholds, they are included, along with nuclear recoil quenching factors~\cite{TRETYAK201040}, as they can produce in shifts in the reconstructed energy. The predicted energy distributions in \naive{} (with arbitrary normalization) from simulation are shown in Fig.~\ref{fig:Sim_Energy}. 

The neutrino flux normalization was parameterized as a function of proton power and energy using a simulation from Ref.~\cite{fluxPaper}, with a 10\% normalization uncertainty. The spectral shape of electron neutrinos from muon decay is well known~\cite{pdg}. Using the MARLEY cross sections, the signal expectations over the 22.8-GWhr exposure for events with visible energy between 10 and 55\,MeV are $\sim$1,320 CC events on \iodine{} and $\sim$61 CC events from \sodium{} and \iron{}. The expected rates for the prompt neutron background were taken from Ref.~\cite{COHERENT:2022eoh}.

Prior to unblinding, the neutron rate and timing shape were studied in low-energy (7--8\,MeV) and high-energy (55--100\,MeV) sidebands to validate the BRN simulation. A one-dimensional binned-timing fit was performed on both of these samples, allowing the neutron normalization and arrival time to float. Additionally, to accommodate differences in time of flight for neutrons of different energy, the timing width was allowed to float by convolving the POT timing distribution with a Gaussian of parameterized width. The neutron rate at the \naive{} location was higher than expected by factor of 6.3 and 9.8 in the respective low- and high-energy sidebands~\cite{COHERENT:2022eoh}. Neutrons in the high-energy sideband were observed 53$\pm6$\,ns earlier than those in the low-energy sideband.  Furthermore, while the high-energy sideband data were consistent with the POT pulse width, the low-energy data preferred a 33$\pm9$~ns additional time broadening. These differences are not understood, but we suspect they are due to lower-energy secondary interactions that can be produced by higher-energy neutrons. Because of these differences, the neutron amplitude, timing offset, and additional width were allowed to float independently in every energy bin.  Thus, we made no prior assumption on the neutron energy or timing distribution.



\textit{Results and discussion.}
The data analysis was blinded with all choices of event selection and fitting finalized before the data were analyzed. Beam events were selected, and the observed energy and timing were reconstructed. Event timing is the principal discriminator between CC and prompt neutron background as $\nu_e$ CC events are delayed from the POT onset.


To mitigate uncertainties from the model of the prompt neutron energy spectrum, the total \nueICC{} cross section was determined by a binned 1D timing fit.  Events with observed energy below 10\,MeV were excluded from the fit to reject neutron capture events (most intense between 4.5 and 6.8\,MeV~\cite{Schaller1971,Islam1990}) which are delayed at a timescale of several microseconds. Additionally, events above 55\,MeV were removed, as these events are beyond the \piDAR{} end-point energy. 


The systematic uncertainty on the \iodine{} CC normalization was 11.4$\%$, dominated by the neutrino flux uncertainty of 10$\%$.  This includes a 5.1$\%$ uncertainty on the fraction of CC events that are rejected by the muon veto panels--this uncertainty originates from the spatial dependence on the threshold within the veto, and its impact was studied through simulation.  We also include uncertainties on the calibration, energy resolution, and neutrino interaction modeling which contribute each $<1\%$.


The steady-state background prediction (predominantly cosmic rays that did not trigger the veto system) was measured \textit{in-situ} with out-of-beam-window data, with negligible uncertainty on its normalization. The prompt neutron flux, arrival time, and timing width were allowed to float without any prior constraint.  We included a $\pm100\%$ uncertainty on the normalization of background \sodium{} and \iron{} CC due to large cross-section uncertainty.  This introduces an uncertainty on the \iodine{} event rate by 31 and 30 events for \sodium{} and \iron{}, respectively.  

\begin{figure}[!bt]
\centering
\includegraphics[width=0.49\textwidth]{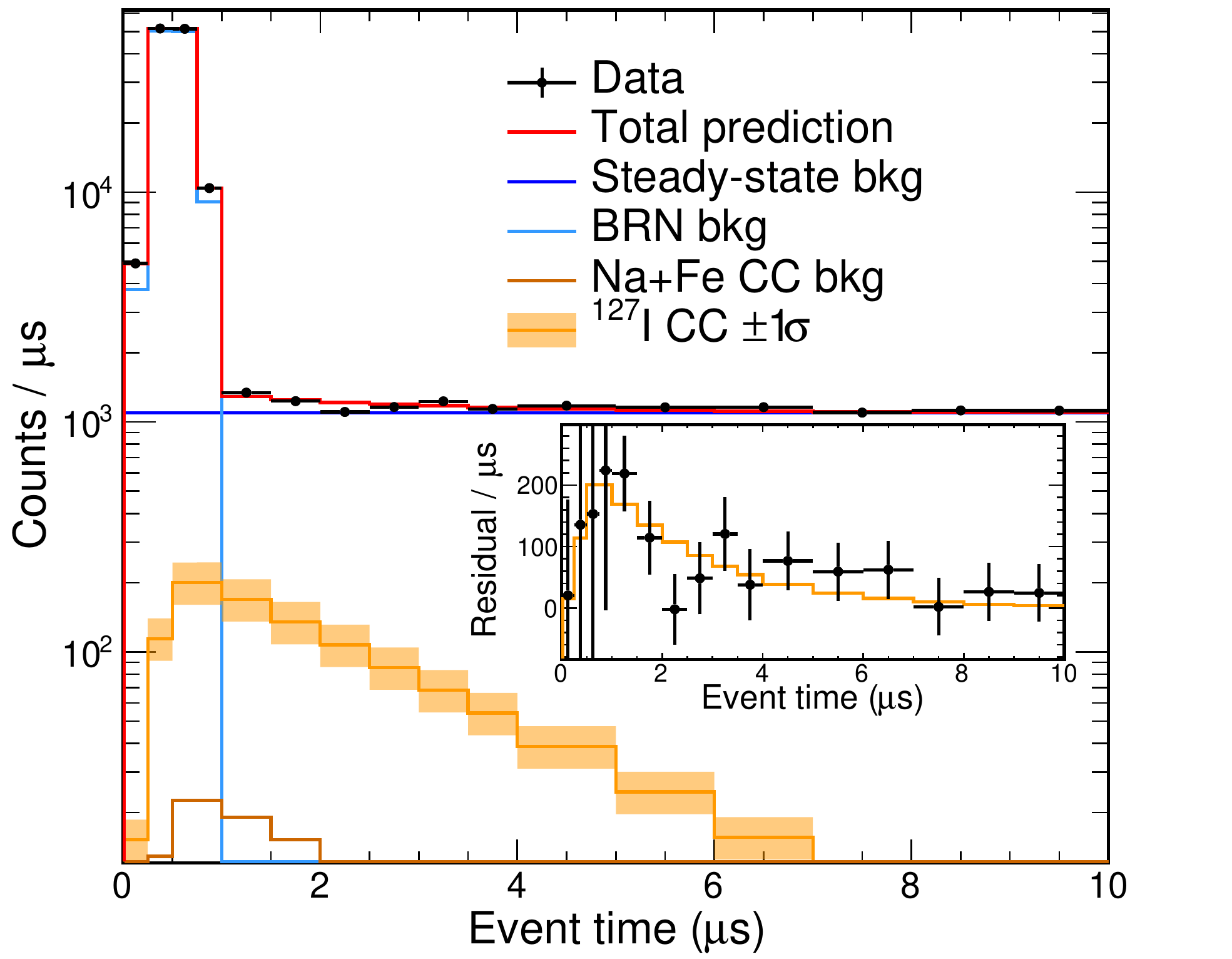}
\caption{The observed \naive{} timing spectrum compared to the total prediction, with an inset showing the background-subtracted timing distribution.  The CC spectrum with 1$\sigma$ uncertainty from the 1D fit is also shown along with the nominal MARLEY prediction and steady-state, BRN, and Na+Fe CC backgrounds.}
\label{fig:SpectrumTiming}
\end{figure}

After selection, we calculated the likelihood curve as a function of the number of \iodine{} CC events while profiling systematic uncertainties.  With this, we find a best fit and 1$\sigma$ range of $541^{+121}_{-108}$, corresponding to a flux-averaged cross section of $(9.2^{+2.1}_{-1.8})\times10^{-40}$~cm$^2$ and inconsistent with the background-only hypothesis at 5.8~$\sigma$.  The best-fit timing spectrum is shown in Fig.~\ref{fig:SpectrumTiming} and has a $\chi^2$ per degree of freedom of 13.1/11.  The best-fit normalization is only 40.9$\%$ of the MARLEY expectation, similar to the NIN suppression COHERENT has observed in lead~\cite{COHERENT:2022eoh}. While disparate from MARLEY's prediction, the measured inclusive cross section is closer to predictions in Ref.~\cite{kosmas1996} $(7.3 \times 10^{-40}\mbox{ cm}^2)$ and Ref.~\cite{athar2006} $(12.5 \times 10^{-40}\mbox{ cm}^2)$. MARLEY's prediction is generated using an assumption of an unquenched $g_A$ (see supplemental material~\cite{naiveSupMat}), and, while the MARLEY model neglects forbidden transitions, we set an upper bound on the suppression of the GT interaction strength for \nueICC{} scattering with \piDAR{} neutrinos of $\leq$ 0.59, corresponding to $g_{A,\mbox{eff}} \leq 0.97$. This upper limit is derived from the lower 1$\sigma$ uncertainty on the GT matrix elements in Ref.~\cite{I_react} and the upper 1$\sigma$ uncertainty on the measured cross section.



We investigate the exclusive channel by considering the energy dependence of \naive{} data due to differences in nuclear binding energy to compare to the measurement of \iodine{}$(\nu_e,e^-)$\xenon{} at LAMPF~\cite{Distel:2002ch} which determined a cross section of $[2.84\pm0.91\mbox{ (stat)}\pm0.25\mbox{ (syst)}]\times10^{-40}$~cm$^2$.  The $1n$ emission threshold is $\sim7$\,MeV larger than that of the $0n$ channel, and much larger than detector resolution. After applying detector resolution, the $40--50$\,MeV range is sensitive to $0n$ events, while the $\geq1n$ channel is kinematically forbidden.  

We performed a 2D fit in time and energy to constrain the $0n$ and $\geq1n$ event normalizations with \naive{} data.  As previously described, the prompt neutron energy distribution is known to be mismodeled, and thus the neutron normalization and timing is floated in each energy bin independently.  The 2D fit is susceptible to uncertainties on the MARLEY predictions for the shape of the observed energy distribution; however the relative fraction of $0n$ and $\geq1n$ events dominate spectral distortions. Thus, we introduce only two uncorrelated fit parameters: the amplitude of $0n$ and $\geq1n$ events. Additionally, the bias introduced on the cross sections is much smaller than statistical uncertainty in this measurement.  These effects will be included in a future measurement from the NaI[Tl] Neutrino Experiment TonnE-scale (\naivete{}) 3.5-tonne detector, currently being deployed at the SNS, which will dramatically reduce statistical errors. 


\begin{figure}[!bt]
\centering
\includegraphics[width=0.49\textwidth]{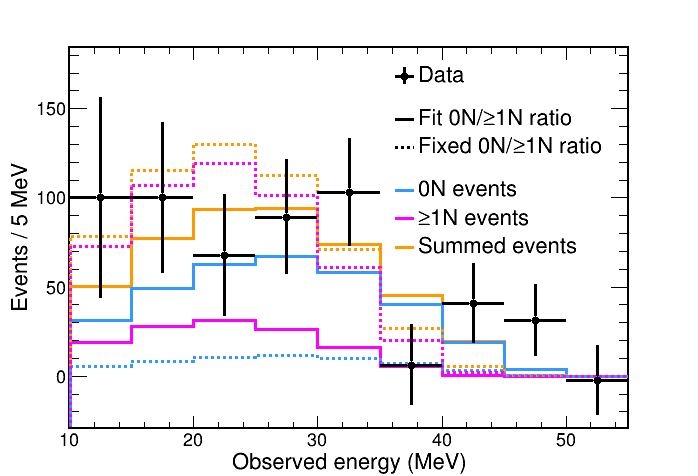}
\caption{The visible energy spectrum of CC events between 10 and 55\,MeV is shown in black, along with the best-fit spectrum from MARLEY (orange) allowing the $\geq1n$ and $0n$ amplitudes to float.}
\label{fig:SpectrumEnergy}
\end{figure}

The resulting energy distribution from the 2D fit is shown in Fig.~\ref{fig:SpectrumEnergy}. The fit agrees well with the model, $\chi^2/d.o.f. = 147.3/139$.  Interestingly, the fit prefers a much higher fraction of $0n$ events (72.3$\%$) than MARLEY predicts (10.6$\%$), though this preference is not strong (a $\Delta\chi^2=3.3$ between the best fit and the MARLEY $0n$ fraction).  Further data collected with the COHERENT \naive{} and \naivete{} detectors can be used to investigate the cross sections of these exclusive channels to tune event generators for CC interactions on heavy nuclei at energies relevant for \piDAR{}, solar, and supernova neutrino measurements.  

\begin{figure}[!bt]
\centering
\includegraphics[width=0.49\textwidth]{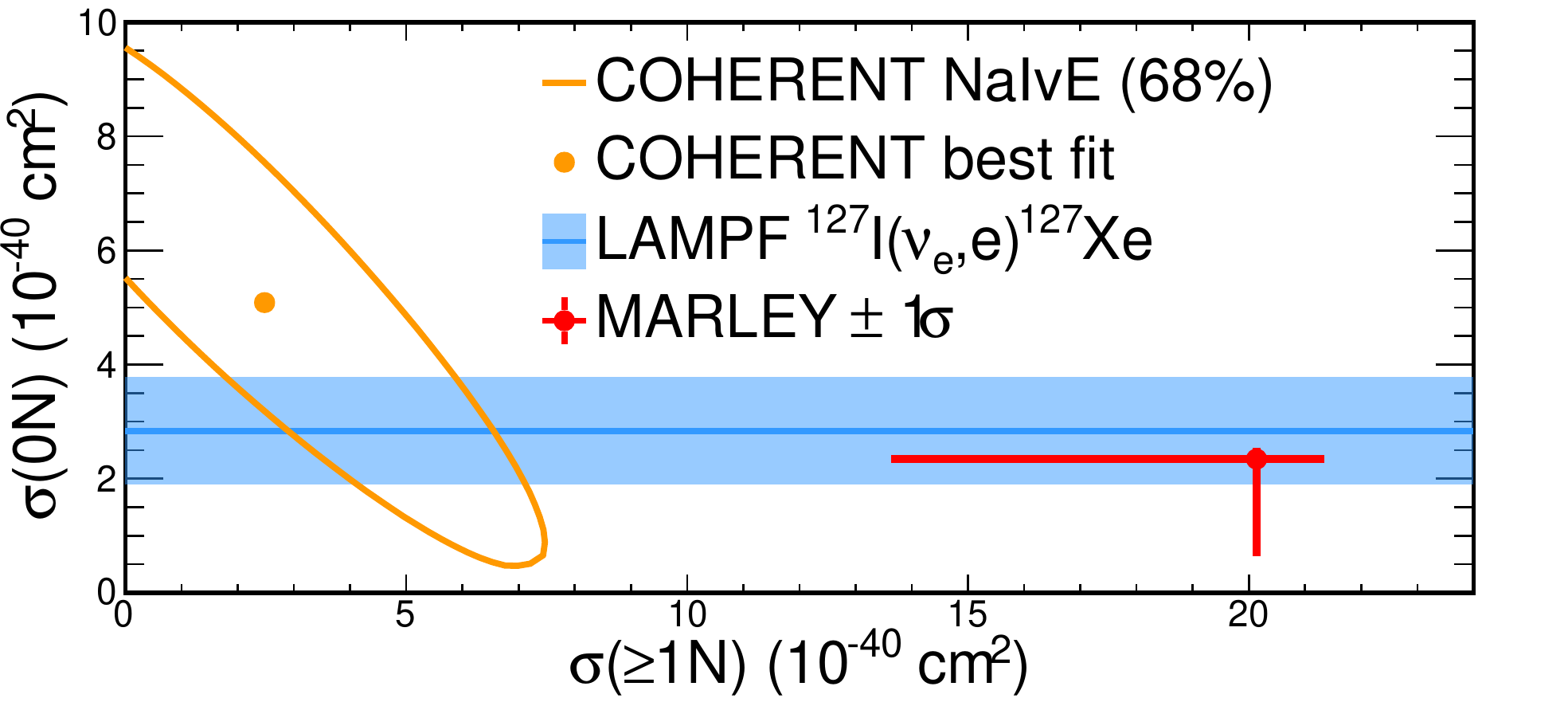}
\caption{Measurement (1$\sigma$) of the \nueICC{} cross section separated into $0n$ and $\geq1n$ channels compared to the MARLEY prediction and Ref.~\cite{Distel:2002ch}, measuring the $0n$ cross section.}
\label{fig:XSec2D}
\end{figure}

From the 2D fit, we derive measurements of the cross sections to the exclusive $0n$ and $\geq1n$ channels simultaneously.  Our measurement is shown in Fig.~\ref{fig:XSec2D}. At 1$\sigma$, the \naive{} data imply $\sigma(0n)=(5.2^{+3.4}_{-3.1})\times10^{-40}$~cm$^2$ after profiling $\sigma(\geq1n)$, consistent with Ref.~\cite{Distel:2002ch} and MARLEY's prediction~\cite{naiveSupMat}, though uncertainties are large due to the $\geq1n$ events present in \naive{}. The determined $1\sigma$ range for $\sigma(\geq1n)$ is $2.2_{-2.2}^{+3.5} \times 10^{-40}\mbox{ cm}^2$ is roughly $10\times$ lower than the MARLEY model, suggesting the suppression in the total rate relative to MARLEY is due to the modeling of the $\geq1n$ channel. Profiles for the exclusive cross-section fits can be found in Supplemental Material~\cite{naiveSupMat}, which includes Refs.~\cite{suzuki1987,talys,taddeucci1987,champagne1989,engel1991,vogel1994,bahcall1996,Na_react,ensdf_iron}. 

\textit{Conclusion:}
COHERENT has measured the inclusive \nueICC{} cross section on \iodine{} between 10 and 55\,MeV to be $(9.2^{+2.1}_{-1.8})\times10^{-40}$~cm$^2$. This measurement is roughly 41\% of the nominal cross section from MARLEY, and to date is the heaviest CC neutrino-nucleus cross section measured in this energy regime. One source of suppression could arise from a quenched value of $g_A$, and while this value cannot be determined due to the lack of forbidden transitions in the MARLEY model, an upper limit of $g_{A,eff} \leq 0.97$ is set. COHERENT's $0n$ emission measurement is consistent with the exclusive channel reported by E-1213, and with that predicted by the MARLEY model. The fit to the $\geq1n$ emission cross section is much smaller than predicted, similar to the suppression observed by COHERENT's previous measurement on lead. The detector continues to collect data, and the future 3.5T \naivete{} detector will improve the current statistical limitations. There are additional efforts within the collaboration to utilize machine-learning approaches on \naive{} data to further improve signal to background. COHERENT's future inelastic detectors will study interactions on ${}^2$H, ${}^{16}$O, ${}^{40}$Ar, and ${}^{232}$Th, significantly increasing the number of neutrino-nucleus interactions studied at these energies.

\textit{Acknowledgments:} The COHERENT Collaboration acknowledges the generous resources provided by the ORNL Spallation Neutron Source, a DOE Office of Science User Facility, and thanks Fermilab for the continuing loan of the CENNS-10 detector. We also thank the Duke physics machine shop for their support in the production of the \naive{} shielding and veto panels. We also acknowledge support from the Alfred~P. Sloan Foundation, the Consortium for Nonproliferation Enabling Capabilities, the National Science Foundation, the Korea National Research Foundation (NRF 2022R1A3B1078756), and the U.S. Department of Energy, Office of Science. Laboratory Directed Research and Development funds from ORNL also supported this project. This work was performed under the auspices of the U.S. Department of Energy by Lawrence Livermore National Laboratory under Contract DE-AC52-07NA27344. This research used the Oak Ridge Leadership Computing Facility, which is a DOE Office of Science User Facility. The work was supported by the Ministry of Science and Higher Education of the Russian Federation, Project ``New Phenomena in Particle Physics and the Early Universe" FSWU-2023-0073.

\bibliography{main}


\section{Supplemental Material: Measurement of electron-neutrino charged-current cross sections on ${}^{127}$I with the COHERENT  NaI$\nu$E detector}
\subsection{Energy Calibrations}
Two prominent peaks were observed at low energies in the NaI[Tl] crystals, originating from ${}^{40}$K (${}^{208}$Tl) decays with an energy of 1.461\,MeV (2.615\,MeV). Each peak was fit with a Gaussian plus a linear background to determine the ADC value corresponding to the peak and energy resolution parameters. Data were calibrated every 12-to-24 hours to track changes in detector gain that can result from temperature fluctuations and PMT aging.

After the low-energy calibration, differences on the order of 5\% were observed in the background muon spectrum of individual NaI[Tl] crystals, believed to originate from non-linear PMT response at higher energies. Michel positrons were used to calibrate in the energy region-of-interest (ROI), 10-to-55\,MeV, relevant for \nueICC{} signals. Michel positrons, produced by stopped cosmic muons decaying within the detector, were selected by searching for events more than 10\,$\mu$s after a muon-tagged event, with the time window set by trace length. The 10-$\mu$s window is sufficiently long to ignore stopped $\mu^-$ capturing on iodine (capture time of 83.4-86.1\,ns~\cite{suzuki1987}) so only $\mu^+$ decays remain. These delayed candidate Michel positrons were required to be contained within a single crystal and to not be coincident with muon veto activity. As a check, these events were plotted and fit with an exponential plus flat background yielding an anti-muon lifetime of 2.201$\pm$0.015\,$\mu$s, in agreement with the experimentally measured lifetime~\cite{pdg}. Positrons were simulated in GEANT4~\cite{geant} with a Michel spectrum and were processed by applying identical cuts as in the data selection to produce a comparable spectrum. The simulations of each channel are fit to the data using a quadratic polynomial with two coefficients fixed to retain the low-energy calibration points. 

\subsection{MARLEY Details}
MARLEY simulates the allowed component of low-energy inelastic neutrino-nucleus distributions. The primary inputs to MARLEY are incident neutrino spectra along with Gamow-Teller, $B(GT^-)$, and Fermi, $B(F)$, strength distributions. Additionally, MARLEY utilizes data from TALYS~\cite{talys} to model discrete nuclear de-excitations. Details on the physics and implementation of MARLEY can be found in Ref.~\cite{gardiner2021_2,gardiner2021simulating}.

Gamow-Teller strength distributions can be obtained from charge-exchange reactions such as $(p,n)$ or (${}^{3}\mbox{He},t$), or can be calculated theoretically. The extraction of Gamow-Teller strengths from $(p,n)$ charge-exchange reactions typically rely on the normalization in Ref.~\cite{taddeucci1987}, in which $\beta$-decay $ft$ values are used. That normalization required an assumption of $g_A = 1.26$, which we have adopted. The definition of the Gamow-Teller strength in $(p,n)$ charge-exchange reactions differs from that used in MARLEY by a factor of $g_A^2$.  We follow the MARLEY convention, $g_A^2=1.26^2$.

The Fermi strength, $B(F)$, is obtained using the Fermi sum rule, $B(F) = g_V^2(N - Z)$, with $g_V=1$ and is located at the isobaric analog state. 

Forbidden transitions are not currently included in MARLEY, and while there have been some studies of non-spin-flip transition strengths for \iodine{}~\cite{champagne1989}, additional measurements and improved theoretical models are needed.

\subsubsection{Charged-Current Predictions for ${}^{127}$I with MARLEY}
For \iodine{}, Gamow-Teller strength was measured via the $(p,n)$ reaction in Ref.~\cite{I_react}. Within that reference, systematic uncertainty is assigned to the measured $B(GT^-)$ strength, with the largest contributors being the normalization for the ratio of unit Gamow-Teller and Fermi cross sections, $E_0$, and the thickness of the scattering target. The uncertainty on the $B(GT^-)$ normalization originates from observation of measurements for odd-odd nuclei resulting in smaller values of $E_0$~\cite{I_react}. The total systematic error given on the sum $B(GT)$ is given within that paper as $\sum{B(GT)}=53.54_{-19.47}^{+3.32}$, but systematic errors are not provided for each GT strength bin. The systematic error is combined with statistical errors in quadrature to produce the given uncertainties. 

The Gamow-Teller strengths used with MARLEY were taken from Tab.~1 within Ref.~\cite{I_react} which provides the strength in 0.5\,MeV bins from 0-19.99 MeV. In preparation of the MARLEY input file, that strength was located at the center of each energy bin. The strength to the lowest excited state in \xenon{} (124.75~keV) is provided within the text Ref.~\cite{I_react}, so it was separated out from the first bin of that table. As mentioned, these strengths were multiplied by $g_A^2=1.26^2$ to convert to the input format required by MARLEY. The statistical uncertainties from Tab.~1 along with the $E_0$ normalization uncertainty were combined in quadrature.

The isobaric analog state (IAS) was measured within Ref.~\cite{I_react} to be centered at 12.68\,MeV. The assigned strength to that state was $B(F)=21$.

As a check of the extraction of the GT strengths from Ref.~\cite{I_react}, Tab.~\ref{tab:solarNu} compares solar neutrino cross section predictions from MARLEY to those within Ref.~\cite{I_react} that used the same Gamow-Teller strengths. While details of how the authors of~\cite{I_react} distributed the Gamow-Teller strength and systematic uncertainties are not known, MARLEY's prediction agrees within uncertainties. 
{\renewcommand{\arraystretch}{1.3}
\begin{table}[!tb]
\begin{center}
\begin{tabular}{ccc}
\hline\hline
Channel & MARLEY & Ref.~\cite{I_react} \\
\hline
${}^7$Be&$1.3_{-0.5}^{+0.2}$&$1.2_{-0.4}^{+0.4}$ \\
${}^8$B&$(5.1_{-1.9}^{+0.5}) \times 10^3$&$(4.3_{-0.6}^{+0.6}) \times 10^3$ \\
\hline
\end{tabular}
\caption{Flux-averaged solar-neutrino cross section predictions from MARLEY compared with those in Ref.~\cite{I_react}, which was the source of the GT data used with MARLEY. All cross sections are in units of ($\times 10^{-45}\mbox{cm}^2$).}
\label{tab:solarNu}
\end{center}
\end{table}
There are other calculations of the \nueICC{} cross section for ${}^8$B solar neutrinos~\cite{haxton1988,I_react,engel1991,vogel1994,lutostansky2021} using other techniques, whose predictions range from $2.2 \times 10^{-42}\mbox{cm}^2$\cite{engel1991} to $8.9 \times 10^{-42}\mbox{cm}^2$\cite{haxton1988}. Using the ${}^8$B solar neutrino flux from Ref.~\cite{bahcall1996}, MARLEY's prediction for the flux-averaged cross section is $5.1_{-1.9}^{+0.5} \times 10^{-42}\mbox{cm}^2$, in agreement with other predictions.

For \piDAR{} \nueICC{} interactions, MARLEY's prediction for the inclusive cross section is $22.5_{-6.5}^{+1.2} \times 10^{-40}\mbox{cm}^2$. This is larger than the majority of existing predictions\cite{kosmas1996,mintz2000,athar2006}, which range from $2.1 \times 10^{-40}$~\cite{engel1994} to $12.5 \times 10^{-40}\mbox{cm}^2$~\cite{athar2006}. We note that with MARLEY's predictions, there are large uncertainties on matrix elements originating from the GT strength measurement, and that quenching of $g_A$ could bring MARLEY's predictions into better agreement with other existing calculations. Further, it is unclear whether there should be quenching in experimentally-measured Gamow-Teller strength distributions. 

MARLEY also predicts exclusive cross sections resulting from \nueICC{} interactions. For exclusive cross sections to bound states of \xenon{}, MARLEY's prediction is $2.5_{-0.9}^{+0.2}\times 10^{-40}\mbox{cm}^2$. This agrees with the measured value from Ref.~\cite{Distel:2002ch} of $2.84\pm0.91\mbox{ (stat)}\pm0.25\mbox{ (syst)} \times 10^{-40}\mbox{cm}^2$. Predictions from MARLEY for exclusive cross sections leading to particle emission can be found in Tab.~\ref{tab:exclusiveChannels}.

{\renewcommand{\arraystretch}{1.3}
\begin{table}[!tb]
\begin{tabular}{cc}
\hline\hline
Channel & Cross section ($\times 10^{-40}\mbox{cm}^2$) \\
\hline
${}^{127}$I$(\nu_e,e^-)$&$22.5_{-6.5}^{+1.2}$\\
${}^{127}$I$(\nu_e,e^-+n){}^{126}$Xe&$18.9_{-5.3}^{+1.0}$\\
${}^{127}$I$(\nu_e,e^-){}^{127}$Xe$_{\mbox{bound}}$&$2.3_{-1.7}^{+0.2}$\\
${}^{127}$I$(\nu_e,e^-+2n){}^{125}$Xe&$0.8_{-0.4}^{+0.1}$\\
${}^{127}$I$(\nu_e,e^-+p){}^{126}$I&$0.5_{-0.2}^{+0.0}$\\
\hline
\end{tabular}
\caption{MARLEY's predictions for inclusive and exclusive \nueICC{} cross sections for \piDAR{} electron neutrinos.}
\label{tab:exclusiveChannels}
\end{table}

Finally, the allowed \nueICC{} cross section prediction on \iodine{} from MARLEY is plotted as a function of incident neutrino energy in Fig.~\ref{fig:MARLEYxs}.

\begin{figure}
  \includegraphics[width=0.5\textwidth]{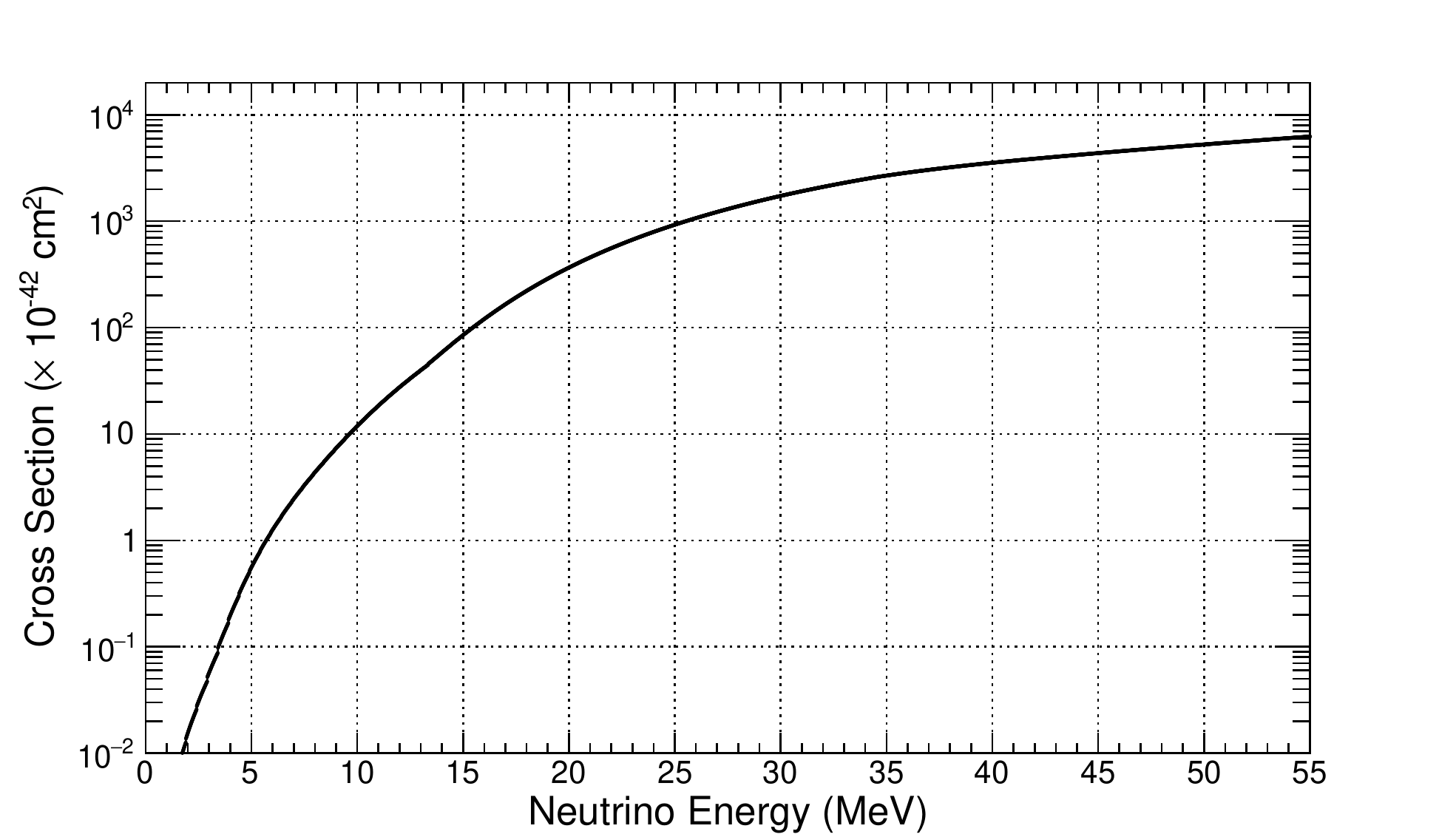}
  \caption{MARLEY's predictions for the electron-neutrino charged-current cross section on \iodine{} using GT strengths from Ref.~\cite{I_react}.}
  \label{fig:MARLEYxs}
\end{figure}

\subsubsection{Other charged-current interactions with MARLEY}
Predictions for backgrounds for charged-current interactions on \sodium{} were generated using Gamow-Teller strength measured in Ref.~\cite{Na_react}. In that reference $({}^{3}\mbox{He},t)$ scattering data were collected and normalized by beta-decay measurements. As a result, the corresponding matrix elements already include the factor of $g_A^2$. The Fermi strength was assumed to be located at the ground state of ${}^{23}$Mg and had a strength of $B(F)=1$. The resulting prediction for the flux-averaged inclusive charged-current cross section on \sodium{} is $0.50 \times 10^{-40}\mbox{ cm}^2$.

For iron, data were taken from Ref.~\cite{Fe_react} using a combination of experimentally measured Gamow-Teller strengths from 0-15.8\,MeV and a RQRPA calculation up to $\sim40$\,MeV excitation energy. The lowest two energy levels were shifted to correspond to Ref.~\cite{ensdf_iron}. Both experimental and theoretical calculations were multiplied by $g_A^2=1.26^2$ to format them for use with MARLEY. The Fermi strength was taken to be located at the IAS at 3.51\,MeV with a value of $B(F)=4$. The resulting prediction for the flux-averaged inclusive charged-current cross section on \iron{} is $2.83 \times 10^{-40}\mbox{cm}^2$, in good agreement with the experimentally measured value from KARMEN of $[2.51 \pm 0.83 \mbox{(stat.)} \pm 0.42 \mbox{(syst.)}] \times 10^{-40}\mbox{ cm}^2$~\cite{maschuw1998}.

\subsection{Determining exclusive cross sections}

In the main text, we describe a 2D fit in recoil time and energy to statistically separate the population of $0n$ and $\geq1n$ events in our selected sample using the difference in $Q$-values for each process.  This is presented as a 2D constraint of both cross sections.  We also calculate the profiled log-likelihood curves for each cross section of these distinct physical processes.  For the $0n$ reaction, the confidence interval determined from the likelihood fit to observed data is $(5.2^{+3.4}_{-3.1})\times10^{-40}$~cm$^2$ which is consistent with the MARLEY prediction.  The $\geq1n$ interaction, however, is determined to be $(2.2^{+3.5}_{-2.2})\times10^{-40}$~cm$^2$, consistent with 0 and roughly an order of magnitude lower than the MARLEY model prediction.  This is an interesting discrepancy and motivates further investigation with the future NaIvETe detector.

\begin{figure}[!ht]
\centering
\includegraphics[width=8.6cm]{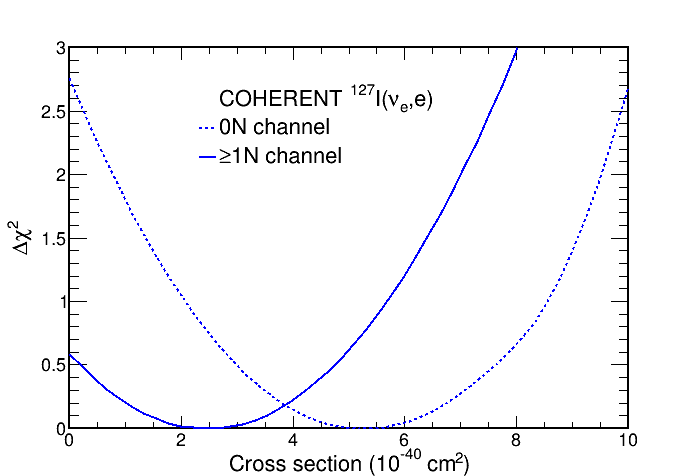}
\caption{The one-dimensional profiled log-likelihood curves for the exclusive $0n$ and $\geq1n$ final states measured with NaIvE data.}
\label{timeTaggedFigure_2}
\end{figure}
\clearpage\newpage

\end{document}